\documentclass[prl,aps,twocolumn,showpacs]{revtex4}
\usepackage{amsmath,amssymb,amsfonts,float}
\usepackage{graphicx,color}  
\usepackage{times,pifont}       
\newcommand{\bolS}{\text{\bf S}}

\newcommand{\bolK}{\text{\bf K}}

\newcommand{\bolk}{\mathbf{k}}
\newcommand{\bolp}{\text{\bf p}}
\newcommand{\bolq}{\mathbf{q}}

\newcommand{\bolQ}{\mathbf{Q}}
\newcommand{\boln}{\mathbf{n}}
\newcommand{\bolr}{\mathbf{r}}

\newcommand{\bolm}{\mathbf{m}}

\newcommand{\VEV}[1]{\langle #1 \rangle}

\newsavebox{\dotdot}
\savebox{\dotdot}[3mm]{\shortstack{\circle*{0.8}\\ \\ \circle*{0.8}}}

\begin{document}
\title{Spontaneous chiral symmetry breaking on a fully polarized frustrated magnet 
at finite temperature
}
\author{Hiroaki~T.~Ueda}
\affiliation{
Okinawa Institute of Science and Technology, Onna-son, Okinawa 904-0412, Japan}
\begin{abstract}
Frustration can introduce more-than-two minima in a spin dispersion relation
even in a fully polarized magnet under high magnetic field. 
We generally discuss, on the fully polarized phase, 
the possibility of the chiral symmetry breaking 
where the number of magnons pumped by finite temperature 
deviates to one side of minima. 
We study this phase by constructing the Ginzburg-Landau energy 
which is controlled by 
the external magnetic field and 
interactions between magnons near the dispersion minima. 
This chirality breaking phase accompanies, not 
a magnetization perpendicular to the external field, 
but the vector chirality $\bolS_\bolm \times \bolS_{\boln}$. 
We also discuss the possibility of 
the chirality breaking phase on LiCuVO$_4$ slightly {\it above} the saturation field.
\end{abstract}
\maketitle
{\it Introduction}-
The concept of a spontaneous symmetry breaking 
has achieved extraordinary successes in the history of 
the condensed matter physics \cite{Goldenfeld}. 
A symmetry breaking phase is characterized by 
what symmetry exists and how it breaks. 
It is well known that, the Heisenberg ferromagnet 
has the SU(2) rotational symmetry in the spin space 
in the Hamiltonian,  and the magnetization accompanies 
the symmetry breaking. 
In addition, various types of symmetry breaking 
phases in magnets have been found. 
One is the spin nematic phase, where not the magnetization 
but the long-range order of
the rank-2 spin tensor breaks 
the rotational symmetry \cite{review_nem}. 
In this letter, we discuss a new type of symmetry breaking, namely, 
the breaking of chiral symmetry in a fully polarized frustrated magnet. 

All spins align in any Heisenberg frustrated magnets by applying 
sufficiently high external field. 
In the fully polarized phase, the spin-flip excitation (magnon) 
can have a nontrivial dispersion relation due to frustration. 
For example, 1-dimensinal (1D) $J_1$-$J_2$ spin chain under high magnetic 
field has two minima in the dispersion relation \cite{Chubukov,HTUandKT,Hikihara,Meisner}. 
Besides, there are several systems realizing two dispersion minima, e.g., the saturated phase in the antiferromagnetic triangular lattice, 
and the $J_1$-$J_2$ models in the square lattice, the cubic lattice, the bcc lattice and so on. 
These two minima are related to each other by the chiral (inversion) symmetry of the lattice. We shall discuss the possibility that magnons at finite temperature,
which has finite density, deviate to one side of minima:
the possibility of the spontaneous chiral symmetry breaking. 
For the concrete discussion, we shall construct the effective Hamiltonian at low temperature and using the mean-field theory. 

There are several compounds which may be viewed as 1D $J_1$-$J_2$ chains 
with interchain couplings \cite{Hase}. 
LiCuVO$_4$ is one of appealing compounds 
which may realize this system \cite{LiCuVO4_1,LiCuVO4_2,NemExp_1,NemExp_2,NemExp_3,NemExp_4,Nawa}: 
near the saturation field, 
the magnetization curve shows a signature of a nontrivial phase \cite{NemExp_3}. 
The trial-wavefunction treatment suggests that 
the anomalous behavior near the saturation field 
is the evidence of a spin-nematic phase \cite{Zhitomirsk-Tsunetsugu}. 
However, microscopic understanding of this anomaly 
in the magnetization curve is still controversial 
both theoretically \cite{HTUandKT2} and experimentally \cite{Nawa2}. 
We shall also discuss the possibility that this anomaly implies the chiral symmetry breaking phase.

{\it Effective Hamiltonian on a saturated phase}-
Let us consider a spin-1/2 Heisenberg model
with generic interactions in a magnetic field:
\begin{equation}
\mathcal{H}=\sum_{\langle i,j\rangle}J_{ij}\,{\bf S}_i{\cdot}{\bf S}_j 
+ H \sum_{j}S^{z}_{j}\ .
\label{Ch1:SpinHamiltonian}
\end{equation}
For convenience, we adopt the hardcore boson
representation of spin operator:
\begin{equation}
S^z_l=-1/2+a^\dagger_la_l \; , \;\;
S_l^+ =a_l^\dagger \; , \; \;  S_l^- =a_l 
\label{Ch2:hardcoreSpin}
\end{equation}
We obtain the following boson Hamiltonian
\begin{equation}
H = \sum_{q}(\omega (\bolq) - \mu)a^\dagger_{\bolq}a_{\bolq}
+\frac{1}{2N}  \sum_{\bolq,\bolk,\bolk^\prime}  V(\bolq) 
a_{\bolk+\bolq}^\dagger a_{\bolk^\prime-\bolq}^\dagger
a_{\bolk}a_{\bolk^\prime}\ ,
\label{Ch2:Hboson}
\end{equation}
\begin{equation}
\begin{split}
&\epsilon(\bolq)=\sum_{j} \frac{1}{2} J_{ij}\cos \left(
\bolq {\cdot}(\bolr_i-\bolr_j)\right)\ ,\\
&\omega(\bolq) =\epsilon(\bolq)-\epsilon_{\text{min}}\ ,\ \ \ 
\mu= H_{\text{c}} -H \ ,\\
& H_{\text{c}} =\epsilon({\bf 0})-\epsilon_{\text{min}}\ ,\ \ \ 
V(\bolq) =2(\epsilon(\bolq)+U)\ ,
\end{split}
\label{bosonRep}
\end{equation}
where $\epsilon_{\text{min}}$ is the minimum of $\epsilon({\bf q})$ 
and $U({\rightarrow}\infty)$ is the hard-core potential 
respecting $S=1/2$ at each site. 
The external field $H$ controls the energy of the magnon, 
which can be viewed as the chemical potential. 
For $H>H_{\text{c}}$ ($\mu<0$), 
the magnon excitation is gapped, 
and the stable ferromagnetic phase is expected.  
If the magnetic field reduces below $H_{\text{c}}$, 
the gap of magnon closes 
and the single-magnon Bose-Einstein condensation (BEC) may occur, 
which leads to $\VEV{S_l^{-}}=\VEV{a_l}\neq 0$. 
In this letter, we concentrate on this gapped fully polarized phase for $\mu<0$.

\begin{figure}[t]
\begin{center}
\includegraphics[scale=0.3]{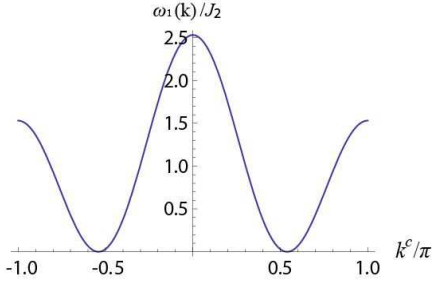}
\caption{
(Color online) The dispersion relation of the 1D $J_1$-$J_2$
chain for $J_1/J_2 = 0.5$ and $J_2 > 0$.
\label{Fig:J1J2_Disp1.eps}}
\end{center}
\end{figure}

As commented before, the dispersion relation 
may have two minima due to frustration. 
For example, the Hamiltonian of the $J_1$-$J_2$ 1D chain 
is given by: 
\begin{equation}
\begin{split}
\mathcal{H}
=\sum_{i}J_1{\bf S}_{i}\cdot {\bf S}_{i+1}+J_2 {\bf S}_{i}\cdot{\bf S}_{i+2}
+ H \sum_{i}S_{i}^z\ ,
\end{split}
\end{equation}
where $J_1<0$, $J_2>0$. For $|J_1/4J_2|<1$, the dispersion minima are
at $q=\pm Q_1$, where $Q_1=\arccos (-J_1/4J_2)$ as shown in Fig.~\ref{Fig:J1J2_Disp1.eps}.
As discussed later, the effect of interchain coupling may be easily included by extending this Hamiltonian.
In the following discussion, for simplicity, 
we assume that the dispersion minima 
are located at $\bolq=\pm\bolQ$ in a 3-dimensional magnet. 

\begin{figure}[H]
\begin{center}
\includegraphics[scale=0.45]{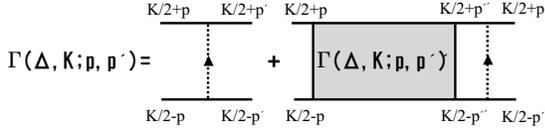}
\caption{Scattering amplitude $\Gamma$ given by the ladder diagram.
\label{Fig:scatteringG}}
\end{center}
\end{figure}

At sufficiently low temperature, 
the underlying physics may be understood by 
magnons near the dispersion minima. 
The important physical quantity is the interaction 
between magnons near the minima, 
which is given by the ladder diagram of Fig.~\ref{Fig:scatteringG}:
\begin{equation}
\begin{split}
&\Gamma(\bolK;\bolp,\bolp^\prime)=V(\bolp^\prime-\bolp)+V(-\bolp^\prime-\bolp)\\
&-\frac{1}{2}\int \frac{d^d p^{\prime\prime}}{(2\pi)^d}\frac{\Gamma(\bolK;\bolp,\bolp^{\prime\prime})(V(\bolp^\prime-\bolp^{\prime\prime})+V(-\bolp^\prime-\bolp^{\prime\prime}))}{\omega(\bolK/2+\bolp^{\prime\prime})+\omega(\bolK/2-\bolp^{\prime\prime})-i0^+}\ ,
\end{split}
\label{Ch2:laddereq}
\end{equation}
where $\bolK$ and $\bolp$ are respectively the center-of-mass momentum of 
the two magnons, and the relative momentum. 
The interaction $\Gamma_1$ between the magnons at the same minimum 
is given by $\Gamma_1=\Gamma(2\bolQ;0,0)/2$. 
The interaction $\Gamma_2$ between magnons at different minima is given by
$\Gamma_2=\Gamma(0;\bolQ,\bolQ)$. 
$\Gamma_1$ and $\Gamma_2$ can be exactly solved \cite{HTUandKT,Batyev,Nikuni-Shiba-2}. 

The important observation is that,
for $\mu<0$ (i.e., $H>H_c$), 
$\Gamma_{1,2}$ are independent of $\mu$. 
Hence, if we write the magnon operators near $\pm Q$ as $u_\bolk=a_{\bolk+\bolQ}$ 
and $v_\bolk=a_{\bolk-\bolQ}$ for small $\bolk$, 
the following effective Hamiltonian may be valid for $\mu<0$:
\begin{equation}
\begin{split}
&H=\sum_{\bolk} (\sum_{i=a,b,c}\frac{k^2_i}{2m_i}-\mu)
(u^\dagger_\bolk u_\bolk+v^\dagger_\bolk v_\bolk)\\
&+\frac{1}{N}\sum_{\bolk_1,\bolk_2,\bolq}\frac{\Gamma_1}{2}
(u^\dagger_{\bolk_1+\bolq} u_{\bolk_2-\bolq}^\dagger u_{\bolk_1} u_{\bolk_2}
+v^\dagger_{\bolk_1+\bolq} v_{\bolk_2-\bolq}^\dagger v_{\bolk_1} 
v_{\bolk_2})\\
&+\Gamma_2 u^\dagger_{\bolk_1+\bolq}u_{\bolk_1} v_{\bolk_2-\bolq}^\dagger 
v_{\bolk_2}+\cdots\ ,
\end{split}
\end{equation}
where the mass term $m$ is obtained by expanding $\omega(\bolk)$ near 
the minima up to the quadratic term (we assume $m$ is the same for 
two minima). We also approximated that the interaction 
between magnons {\it near} minima 
are given by $\Gamma_{1,2}$. 
We drop the higher order interaction term, which may not be important 
when the density of magnon is small. 
We concentrate on the case that $\Gamma_1>0$;
otherwise, the higher order interaction terms
(represented by dots) are needed to suppress the density of magnons. 

Even in a fully polarized phase for $\mu<0$, 
there are finite density of magnons at finite temperature. 
Naively we expect that, 
if the interaction $\Gamma_2$ between $u$ and $v$ bosons is repulsive and sufficiently large relative to $\Gamma_1$, 
magnons favor decline to one side of minima in the energetic viewpoint, 
which leads to the chiral symmetry breaking. 

Before going into the details, let us 
comment on the relevance between our scenario and the previous studies. 
The concept of the deviation of two-component bosons itself 
is not specific to our case. 
For example, in our setting, 
slightly {\it below} the saturation field for $\mu>0$, 
magnon BEC may occur. If $0<\Gamma_1<\Gamma_2$, 
there appears the difference of the densities of condensed magnons 
for $\bolk=\pm\bolQ$, 
which accompanies the chirality breaking (cone phase) \cite{HTUandKT}. 
In addition, the frustrated 1D-spin tube at zero temperature 
under high magnetic field  
can be described by many-component bosonic model, 
and the deviation of the densities can also lead to the chirality breaking \cite{Sato1}. 
The recent numerical simulation confirms that 
the spontaneous population imbalance of two-component bosons 
actually occurs in the 1D system at zero temperature \cite{Sato2}. 

Next, we study the possibility of our scenario 
by using the mean-field approximation. 
In advance, we note that
the chirality breaking may be the same university class as the Ising 
transition. Hence, even if it seems to appear in any dimensions 
within the mean-field treatment, 
it may occur only in 2D- or 3D magnetic systems 
at finite temperature.

\begin{figure}[t]
\begin{center}
\includegraphics[scale=0.25]{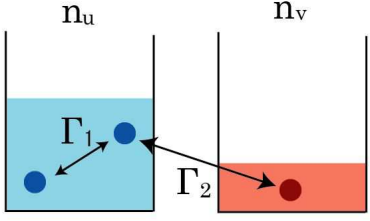}
\caption{(color online) Schematic figure for the chirality breaking phase.
If the interaction $\Gamma_1$ between the same ones is 
sufficiently smaller than the $\Gamma_2$ between the different ones, 
the densities of bosons can become different: $n_u\neq n_v$ ($n_s\neq0 $).
\label{Fig:cbp}}
\end{center}
\end{figure}

{\it Mean field treatment-} 
At finite temperature, the density of magnons is finite:
$n_u=(1/N)\sum_\bolk\VEV{u^\dagger_\bolk u_\bolk}\neq 0,\ n_v=(1/N)\sum_\bolk\VEV{v^\dagger_\bolk v_\bolk}\neq 0$. If we neglect the term $((1/N)\sum_\bolk 
u^\dagger_\bolk u_\bolk-n_u)^2$ and $((1/N)\sum_\bolk v^\dagger_\bolk v_\bolk-n_v)^2$, we obtain the mean-field Hamiltonian:
\begin{equation}
H_\text{mean}=\sum_\bolk (\epsilon_\bolk^u
u^\dagger_\bolk u_\bolk
+\epsilon_\bolk^v v^\dagger_\bolk v_\bolk)
-\Gamma_1N(n_u^2+n_v^2)-\Gamma_2 Nn_u n_v\ ,
\end{equation}
where
\begin{equation}
\begin{split}
\epsilon_\bolk^u&
=\sum_{i=x,y,z}\frac{k^2_i}{2m_i}-\mu+\Gamma_t n_t+\Gamma_s n_s\ ,\\
\epsilon_\bolk^v&
=\sum_{i=x,y,z}\frac{k^2_i}{2m_i}-\mu+\Gamma_t n_t-\Gamma_s n_s\ ,\\
n_t&=n_u+n_v\ ,\ n_s=n_u-n_v\ ,\ \\
\Gamma_t&=\Gamma_1+\frac{\Gamma_2}{2}
\ ,\ 
\Gamma_s=\Gamma_1-\frac{\Gamma_2}{2}\ ,
\end{split}
\end{equation}
$n_s$ can be viewed as the order parameter of the chirality breaking phase.
Then, the free energy at finite temperature is given by
\begin{equation}
\begin{split}
\frac{F}{N}
&=\frac{1}{N}\sum_\bolk \frac{1}{\beta}
(\log (1-e^{-\beta \epsilon_\bolk^u})+\log (1-e^{-\beta \epsilon_\bolk^v}))\\
&-\frac{\Gamma_t}{2}n_t^2-\frac{\Gamma_s}{2} n_s^2\ .
\end{split}
\end{equation}
$\frac{\partial F}{\partial n_t}=0$ and $\frac{\partial F}{\partial n_s}=0$ produces the self-consistent equations
\begin{subequations}
\begin{align}
n_t=\frac{1}{N}\sum_\bolk (g_b(\epsilon_\bolk^u)+g_b(\epsilon_\bolk^v))\ ,
\label{nt_selfeq}\\ 
n_s=\frac{1}{N}\sum_\bolk (g_b(\epsilon_\bolk^u)-g_b(\epsilon_\bolk^v))\ ,\label{n_s_mean}
\end{align}
\label{mean_eq}
\end{subequations}
where $g_b(\epsilon)=1/(e^{\beta \epsilon}-1)$.
Note that, if $\Gamma_s>0$ and $n_s >0$ ($n_s<0$), the right-hand side of (\ref{n_s_mean}) becomes negative (positive): there is no solution of $n_s\neq0$ 
for $\Gamma_s>0$ and the chirality is never broken. 
Hence, in the following discussion we always assume $\Gamma_s<0$.
Later, we shall see the concrete model of $n_s\neq 0$. 
The schematic figure of this chirality breaking is shown in Fig.~\ref{Fig:cbp}.

When this phase appears, 
the z-component of the vector chirality 
$(\bolS_l\times \bolS_m)^z$ has a finite value and 
can be viewed as the order parameter. 
On the other hand, the magnetization 
does not display nontrivial behaviors in this phase: 
$\VEV{S^z_l}=$const, $\VEV{S^x_l}=\VEV{S^y_l}=0$.
Hence, the order parameter of this phase 
is the same as the chiral phase 
at zero temperature found in the pure-1D $J_1$-$J_2$ 
chain \cite{1DJ1J2_chiral}.

Before seeing the concrete example, let us discuss 
the general aspect of the second-order-phase transition 
by using the Landau expansion.

{\it Landau free energy with respect to $n_s$-}
We assume that eq.~(\ref{nt_selfeq}) is satisfied. 
Then, the Landau free energy for $n_s$ is given by
\begin{equation}
L=\frac{d^2 F}{d n_s^2}|_{n_s=0}n_s^2+O(n_s^4)\ .
\end{equation}
Hence, if $\frac{d^2 F}{d n_s^2}|_{n_s=0}<0$, 
$n_s\neq 0$ and the chirality breaking phase is expected. 
At $\frac{d^2 F}{d n_s^2}|_{n_s=0}=0$, the second order phase 
transition may occur.
Explicitly, 
\begin{equation}
\frac{d F}{d n_s}=\frac{\partial F}{\partial n_s}
+\frac{\partial F}{\partial n_t}\frac{d n_t}{d n_s}=
(\frac{1}{N}\sum_\bolk (g_b(\epsilon_\bolk^u)-g_b(\epsilon_\bolk^v)))\Gamma_s-\Gamma_s n_s\ ,
\end{equation}
where we use $\frac{\partial F}{\partial n_t}=0$. Thus
\begin{equation}
\begin{split}
\frac{d^2 F}{d n_s^2}|_{n_s=0}&=(\frac{2}{N}\sum_\bolk g_b^{\prime}(\epsilon_\bolk))\Gamma_s^2-\Gamma_s\\
&=\frac{1}{N}\sum \frac{-2\beta\Gamma_s^2 e^{\beta \epsilon_k}}{(e^{\beta \epsilon_k}-1)^2}-\Gamma_s\ ,
\end{split}
\label{second_c}
\end{equation}
where we use $\frac{d n_t}{d n_s}|_{n_s=0}=0$ and $\epsilon_\bolk=\frac{\bolk^2}{2m}-\mu+\Gamma_t n_t$. 

\begin{figure}[t]
\begin{center}
\includegraphics[scale=0.3]{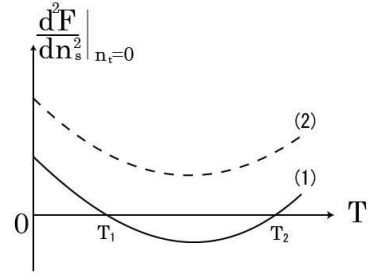}
\caption{
Re-entrant behavior of the chirality braking phase within the Landau 
theory. 
Schematically $\frac{d^2 F}{d n_s^2}|_{n_s=0}$ for fixed $\mu$ is shown. 
For $T\rightarrow 0$ and $\infty$, eqs. (\ref{t0}) 
and (\ref{tinf}) give $\frac{d^2 F}{d n_s^2}|_{n_s=0}>0$.
Case (1): the chiral symmetry breaking phase appears between 
$T_1$ and $T_2$. 
Case (2): the chiral symmetry is not broken.
Re-entrant is because a critical density of thermally excited 
magnons may be needed to stabilizes the state at low temperature. 
For high temperature, the entropy may become dominant.
\label{Fig:scem}}
\end{center}
\end{figure}

For the low temperature limit $\beta\rightarrow \infty$, we obtain
\begin{equation}
\frac{d^2 F}{d^2 n_s}|_{n_s=0}\approx_{\beta\rightarrow \infty} -\Gamma_s>0\ ,
\label{t0}
\end{equation}
which means a non-chirality breaking phase.
This may be because the effect of interactions is irrelevant 
in the dilute limit. 
For the large temperature limit $\beta\rightarrow 0$, 
\begin{equation}
\begin{split}
&\frac{d^2 F}{d n_s^2}|_{n_s=0}\approx_{\beta\rightarrow 0} -\Gamma_s(\frac{\Gamma_s}{\Gamma_t}+1)>0\ ,\label{tinf}\\
&n_t\approx\sqrt{\frac{2}{\beta\Gamma_t}}\ ,
\end{split}
\end{equation}
where we use $\Gamma_t>|\Gamma_s|$ since we assumed that $\Gamma_1>0$ and 
$\Gamma_s<0$ ($2\Gamma_2>\Gamma_1>0$).
For large and small $\beta$ limit, $\frac{d^2 F}{d n_s^2}|_{n_s=0}>0$: 
the chirality breaking phase does not appear. 
Hence, it can appear only between an appropriate temperature region
as illustrated in Fig.~\ref{Fig:scem}.  

The other limit gives us the information of the qualitative behavior. 
For the large $-\mu$ limit, we obtain
\begin{equation}
\frac{d^2 F}{d^2 n_s}|_{n_s=0}\approx_{\mu\rightarrow -\infty}-\Gamma_s>0\ .
\end{equation}
Hence, if the chirality breaking phase appears under any external field, 
this phase disappears by increasing the external field further enough. 

\begin{figure}[H]
\begin{center}
\includegraphics[scale=0.3]{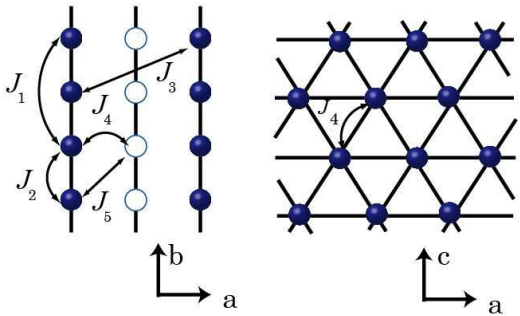}
\caption{The lattice structure and the exchange interactions 
of LiCuVO$_4$\cite{LiCuVO4_2}. Dots represent $S=1/2$ spins.
\label{Fig:LiCuVO4}}
\end{center}
\end{figure}

{\it Concrete calculation-}
Finally, let us concretely calculate eq.~(\ref{mean_eq}). 
We use the input parameters which have a relevance to LiCuVO$_4$ 
as illustrated in Fig.~\ref{Fig:LiCuVO4}.
The mass parameter is directly derived as $m^{(1)}_a=3.7$, $m^{(1)}_b=6.7$, and $m^{(1)}_c=0.068$ (meV$^{-1}$), where we use $J_1=-1.6$, $J_2=3.8$, $J_3=0.4$, $J_4=-0.015$, $J_5=0.08$ (meV) following Ref.~\onlinecite{LiCuVO4_2}.

\begin{figure}[bt]
\begin{center}
\includegraphics[scale=0.6]{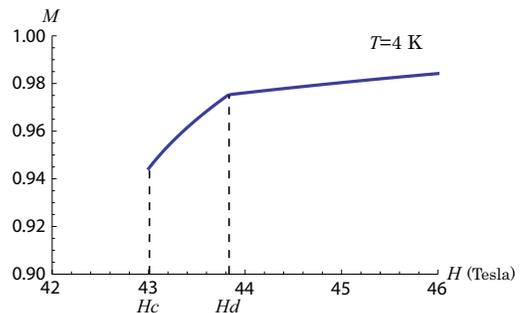}
\caption{The magnetization curve for $\Gamma_1=0.5$, 
$\Gamma_2=50$ (meV) at $T=4$K. $M=\VEV{S_l^z}/S$ for any $l$. 
$H_c=43$ (Tesla) is the saturation field of the single magnon given by 
substituting the 3-dimensional couplings of LiCuVO$_4$ used 
in this letter from Ref.~\cite{LiCuVO4_2} into eq.~(\ref{bosonRep}), 
where the $g$-factor of $2.3$ in the c-direction \cite{NemExp_3} is assumed. 
From $H=H_c$ to $H_d$, the chirality breaking phase appears. 
At $H=H_d$, the second order phase transition occurs. 
For $H>H_d$, the usual fully polarized phase appears.  
\label{Fig:magCB}}
\end{center}
\end{figure}

By solving eq.~(\ref{Ch2:laddereq}), we obtain 
$\Gamma_1=-1.8$, $\Gamma_2=2.9$ (meV). 
Naively, the attractive $\Gamma_1$ interaction and the repulsive $\Gamma_2$ 
interaction may favor the chirality breaking. However, we cannot treat the negative $\Gamma_1$ 
within the previously discussed mean-field theory 
due to the lack of the value of a sextic interaction term. 
Hence, in the following, we use only the mass term of LiCuVO$_4$,
and freely assume various $\Gamma_{1,2}$ of positive $\Gamma_1$. 
In our-assumed-mass parameter, 2-dimensionality may hold below $T\approx 5$K, 
and we focus on the case for $T< 5$K. 
For simplicity, in the numerical calculation, 
the cutoff of $\bolk$ was taken nearly infinity 
since the integrant in eq.~(\ref{mean_eq}) 
becomes negligible for large $k$.
If we assume $\Gamma_1=0.5$, the chirality breaking phase 
does not appear for $T<5$K unless $\Gamma_2>30$. 
If we substitute $\Gamma_1=0.5$ and $\Gamma_2=50$ by hand, 
the chirality breaking phase appears for $T>1.5$K at $\mu=0$.
At $T=4$K, the magnetization curve is shown in Fig.~\ref{Fig:magCB}, 
where $H_c$ is assumed as that of LiCuVO$_4$ given by eq.~(\ref{bosonRep}) \cite{CommentRef1}. 
The chirality breaking phase appears slightly above the saturation field. 
Fig.~\ref{Fig:magCB} qualitatively captures the properties of the anomalous 
behavior near the saturation field in LiCuVO$_4$ \cite{comment22}.

{\it Conclusion and discussion-}
We have suggested the chirality breaking phase 
in the fully saturated phase with two (or more) dispersion minima in frustrated magnets. 
If the interaction between magnons at different minima is repulsive 
and strong enough, the magnons pumped at finite temperature may 
deviate to one side of minima. 
We argued that this phase can exist, if some conditions are met,
by using the effective Hamiltonian and the mean-field 
calculation. 
The order parameter of this phase 
is the z-component of the vector chirality 
$(\bolS_l\times \bolS_m)^z$, 
while the magnetization is usual:
$\VEV{S^z_l}=$const, $\VEV{S^x_l}=\VEV{S^y_l}=0$.
From the Landau expansion, we found that 
this phase is re-entrant and does not appear at the zero-temperature 
and high-temperature limit. 
This may be because, for the zero-temperature limit, 
the density of magnon is small 
and the effect of the interaction is negligible, 
and, for the high-temperature limit, 
the entropy dominates the physics. 
We also found that, for large $-\mu$ limit, the chirality is not 
broken, since the density of magnons becomes small. 
These qualitative behaviors imply: 
(i) this phase can appear for an appropriate temperature region; 
(ii) this phase favors small $|\mu|$ (slightly above the saturation field), 
and increment of $-\mu$, which lowers the 
density of magnons, destabilizes this phase. 
We have also suggested that, the later qualitative behavior (ii)
is consistent with the properties of the anomalous behavior 
of the magnetization curve found in LiCuVO$_4$ \cite{NemExp_3}.
Hence, if the experimentally observed anomalous behavior near saturation 
field actually implies the existence of a nontrivial phase \cite{NemExp_3}, 
and if the phase is not a spin-nematic \cite{HTUandKT2}, 
it may be important 
to experimentally pursue the possibility of the chiral symmetry breaking 
slightly {\it above} the saturation field in LiCuVO$_4$.



{\it Acknowledgments-}
We thank T. Momoi, Y. Sasaki, M. Sato, M. Takigawa, G. Tatara, 
and K. Totsuka for helpful discussions.
We are also grateful to N. Shannon for 
useful discussions and careful reading of this letter. 
This work was supported by JSPS KAKENHI Grant Number 26800209.



\begin{thebibliography}{101}
\bibitem{Goldenfeld}
N.~Goldenfeld, 
{\it Lectures on phase transitions and the renormalization group}
(Westview Press 1992).
\bibitem{review_nem}
K. Penc and A. Lauchli, 
{\it Introduction to frustrated magnetism, chapter 13} 
(Springer-Verlag Berlin Heidelberg
2011)
\bibitem{Chubukov}
A.~V.~Chubukov, Phys.\ Rev.\ B {\bf 44}, 4693 (1991).
\bibitem{HTUandKT}
H.~T.~Ueda and K.~Totsuka, 
Phys. Rev. B {\bf 80}, 014417 (2009).
\bibitem{Hikihara}
T.~Hikihara, L.~Kecke, T.~Momoi, and A.~Furusaki, Phys.\ Rev.\ B
{\bf 78}, 144404 (2008).
\bibitem{Meisner}
F.~Heidrich-Meisner, I.~P.~McCulloch, and A.~K.~Kolezhuk, 
Phys. Rev. B {\bf 80}, 144417 (2009).
\bibitem{Hase}
M.~Hase, {\it et. al.}, 
Phys. Rev. B {\bf 70}, 104426 (2004).
\bibitem{LiCuVO4_1}
B.~J.~Gibson, {\it et. al.}
Physica\ B {\bf 350}, e253 (2004).
\bibitem{LiCuVO4_2}
M.~Enderle, {\it et. al.}
Europhys.\ Lett.\ {\bf 70}, 237 (2005).
\bibitem{NemExp_1}
M.~G.~Banks, {\it et. al.}
J.~Phys.:~Condens. Matt. {\bf 19}, 145227 (2007).
\bibitem{NemExp_2}
N.~B\"{u}ttgen, {\it et. al.}
Phys. Rev. B {\bf 76}, 014440 (2007).
\bibitem{NemExp_3}
L.~E.~Svistov, {\it et. al.}
JETP letters, {\bf 93}, 21 (2011).
\bibitem{NemExp_4}
M.~Mourigal, {\it et. al.}
Phys. Rev. Lett. {\bf 109}, 027203 (2012).
\bibitem{Nawa}
K.~Nawa, {\it et.al.}
J. Phys. Soc. Jpn., {\bf 82}, 094709 (2013).
\bibitem{Zhitomirsk-Tsunetsugu}
M.~E.~Zhitomirsky and H.~Tsunetsugu, 
Europhys. Lett. {\bf 92}, 37001 (2010). 
\bibitem{HTUandKT2}
H.~T.~Ueda and K.~Totsuka, 
arXiv:1406.1960. 
\bibitem{Nawa2}
M. Takigawa, {\it et. al.}, unpublished. 
\bibitem{Batyev}
E.~G.~Batyev and L.~S.~Braginskii, Zh. Eksp. Teor. Fiz. {\bf 87}, 
1361 (1984) [Sov. Phys. JETP {\bf 60}, 781 (1984)]; 
E.~G.~Batyev, Zh. Eksp. Teor. Fiz. {\bf 89}, 
308 (1985) [Sov. Phys. JETP {\bf 62}, 173 (1985)].
\bibitem{Nikuni-Shiba-2}
T.~Nikuni and H.~Shiba, 
J. Phys. Soc. Jpn. {\bf 64}, 3471 (1995). 
\bibitem{Sato1}
M.~Sato and T.~Sakai, 
Phys. Rev. B {\bf 75}, 014411 (2007). 
\bibitem{Sato2}
S.~Takayoshi, M.~Sato, and S.~Furukawa, 
Phys. Rev. A {\bf 81}, 053606 (2010).
\bibitem{1DJ1J2_chiral}
K.~Okunishi, J. Phys. Soc. Jpn. {\bf 77}, 114004 (2008); 
T.~Hikihara, T.~Momoi, A.~Furusaki, and H.~Kawamura, 
Phys. Rev. B {\bf 81}, 224433 (2010).
\bibitem{CommentRef1}
For simplicity, we do not consider the critical field of 
a bound-magnon instability, 
which is known to exist in the parameter 
of LiCuVO$_4$ \cite{Zhitomirsk-Tsunetsugu}.
\bibitem{comment22}
Even if we take into account the fact that 
$\Gamma_2$ is affected twice as much as $\Gamma_1$ 
by change of the original 
scattering amplitude eq.~(\ref{Ch2:laddereq}), 
$\Gamma_1=0.5$ and $\Gamma_2=50$ is not a little different from the original values of LiCuVO$_4$ 
($\Gamma_1=-1.8$ and $\Gamma_2=2.9$), which cannot be treated due to the theoretical difficulty up to now. 
However, this does not exclude the possibility 
of the chirality breaking phase in LiCuVO$_4$.
Our approach is limited within the mean-field treatment, 
which may not be reliable in quasi-1D systems. 
In addition, we did not consider the effect of a bound magnon \cite{Zhitomirsk-Tsunetsugu}, 
the density of which may be of the same order as that of single magnons 
slightly above the saturation field.
Hence, our estimate may be considered as a phenomenological and preliminary one.\end{thebibliography}
\end{document}